\documentclass[]{article}
\usepackage{amsmath}
\usepackage{amsfonts}
\usepackage{amssymb}
\usepackage{graphicx}
\usepackage{amsfonts}
\usepackage[english]{babel}
\usepackage[utf8]{inputenc}
\usepackage{times}
\usepackage[T1]{fontenc}
\usepackage{multirow}

\begin{document}

\title{SU(2) and SU(1,1) Y-Maps in Loop Quantum Gravity}

\author{Leonid Perlov, \\
Department of Physics, University of Massachusetts, Boston, USA\\
leonid.perlov@umb.edu
}

\maketitle

\begin{abstract}
In this paper we first provide the proof of $SU(2)$ Y-Map convergence. Then, by using $SU(1,1)$ LQG simplicity constraints we define $SU(1,1)$ Y-Map  from infinitely differentiable with a compact support functions on $SU(1,1)$ to the functions (not necessarily square integrable) on $SL(2,C)$, and prove its convergence as well. 
\end{abstract}
 
\section{Introduction}
Y-Map takes the central place in the Loop Quantum Gravity as it provides the map from $SU(2)$ spin networks to $SL(2,C)$ spin networks. SU(2)-Y-Map was first introduced in $\cite{Rovelli2010}$, $\cite{RovelliBook2}$ and $\cite{SpinFoam}$.
As it is stated in Rovelli and Vidotto book $\cite{RovelliBook2}$: ''Y-Map is  the core ingredient of the quantum gravity dynamics ... It codes the way $SU(2)$ states transform under $SL(2,C)$ transformations. This in turn, codes the dynamical evolution of the quantum states of space.'' The physical states of quantum gravity are thus the images of  $SU(2)$ spin networks under Y-Maps. Explicitly SU(2)-Y-Map is defined in $\cite{RovelliBook2}$  (7.26, 7.27) as: \\
\begin{equation}
\label{SU(2)Y-Map}
Y_{SU(2)}:   \qquad    L_2[SU(2)]  \rightarrow F[SL(2,C)]
\end{equation}

\begin{equation}
\label{SU(2)Y-Map1}
 \psi(u) = \sum\limits_{jmn}c_{jmn}D^{(j)}_{mn}(u)   \rightarrow   \phi(g) =  \sum\limits_{jmn}c_{jmn} D^{(j, \gamma j, )}_{jm jn}(g) 
\end{equation}
, where $\gamma \in R$ is Immirzi parameter, $u \in SU(2), g \in SL(2,C), j, m, n \in Z$ .\\
Or, since $D^{(j, \gamma j, )}_{jm jn}(g) = 0$ for $m \ne n$, we can rewrite it in a simpler form:\\
\begin{equation}
\label{SU(2)Y-Map1}
 \psi(u) = \sum\limits_{jm}c_{jm}D^{(j)}_{m}(u)   \rightarrow   \phi(g) =  \sum\limits_{jm}c_{jm} D^{(j, \gamma j, )}_{jm jm}(g) 
\end{equation}
It is clear from the above definition that it is very different from the Plancherel formula for the Lorentz group. The latter contains a sum and an integral over all parameters of the principal series matrix coefficients  $D^{(k, \rho)}_{jm, j'n}$, while the $Y_{SU(2)}$-Map takes the Fourier transforms $c_{jmn}$ of the square integrable function on $SU(2)$ and contracts them with the $SL(2,C)$ matrix coefficients $D^{(j, \gamma j)}_{jm jn}(g) $  summing up not over all $SL(2,C)$ parameters $(k, \rho), k\in Z, \rho \in R$, but rather over the selected ones provided by the simplicity constraints spin map: $ (k = j, \rho = \gamma j) $, and thus avoiding an integral, that is present in Plancherel formula.
\\[2ex]
 The simplicity constraints, introduced by John Barrett and Louse Crane in $\cite{BarrettCrain}$ allow us to consider the Quantum Gravity as a 4-dimensional topological model called BF-model plus some constraints on the form of the bivectors used in BF model. Those constraints are called the simplicity constraints. The simplicity constraints make the 4-dim topological model become Einstein's Quantum Gravity. $SU(2)$ simplicity constraints solution is $ (k = j, \rho = \gamma j) $.
For details see $\cite{BarrettCrain}$, $\cite{FlippedVertex}$, $\cite{VertexAmplitude}$ and $\cite{EPRL}$.\\[2ex]
A question to ask is whether the sum on the r.h.s of ($\ref{SU(2)Y-Map1}$) is convergent, in other words if $Y_{SU(2)}$-Map exists.  In Theorem 1 ($SU(2)$ Y-Map Existence Theorem) we prove its convergence by using the essential Lemma 1 for which we also provide the proof. In the Appendix B we show the graph obtained from MPMath Python program $\cite{mpmath}$ demonstrating visually its convergence. We should also mention a different approach to map functions on $SU(2)$  to functions on $SL(2,C)$ that was made in $\cite{LiftingSU2}$. That approach is different from Y-Maps, it was made for $SU(2)$, but not for $SU(1,1)$, and the convergence issue was not considered. 
\\[2ex]
 In this paper we also provide $SU(1,1)$ Y-Map definition and prove that map convergence. While $SU(2)$ Y-Map corresponds to the spacelike ADM foliation of the 4 dimensional spacetime, similarly $SU(1,1)$ Y-Map corresponds to timelike ADM foliation $\cite{ConradyHnybida}$. Recently the timelike foliation has received a deserved attention $\cite{Montesinos}$, $\cite{Noui}$ in an attempt to create a covariant theory in place of a spacelike covariant one.\\[2ex]
SU(1,1)-Y-Map is a map from infinitely differentiable functions on $SU(1,1)$ with a compact support to functions on $SL(2,C)$ provided by the $SU(1,1)$ simplicity constraints solution $\cite{ConradyHnybida}$, which for discrete series is the same as $SU(2)$ one: $ (k = j, \rho = \gamma j) $, while for continuous series is different: $(\rho = -\frac{k}{\gamma}, s = \frac{1}{2}\sqrt{(k^2/\gamma^2-1} ))$\\[2ex]
The paper is organized as follows. In section $\ref{sec:SU(2)-Y-Map}$ we prove $SU(2)$ Y-Map convergence ($SU(2)$ Y-Map Existence Theorem). In section $\ref{sec:SU(1,1)-Y-Map}$ we define $SU(1,1)$ Y-Map and prove its convergence. Section $\ref{sec:Discussion}$ concludes the paper. Appendix B contains the programming code and the graphs demonstrating convergence of the sum of  $SL(2,C)$ matrix coefficients divided by polynomial. This sum is used in Lemma 1 and is essential for $SU(2)$ Y-Map Existence Theorem proof. The graphs have been produced by using MPMath Python program  $\cite{mpmath}$. 
\section{SU(2)-Y-Map}
\label{sec:SU(2)-Y-Map}
We are going to prove that any square integrable function  $\phi(u)$ on $SU(2)$ can be mapped to a function $\psi(g)$ on $SL(2,C)$ (not necessary square integrable), by using the solution of the $SU(2)$ simplicity constraints $\cite{EPRL}$: ($k = j, \rho = \gamma j, j \in Z, \gamma \in C$) in  the following manner:
\begin{equation}
\label{SU(2)YMap}
\phi(u) \rightarrow \psi(g) = \sum \limits_{j=|p|}^{\infty}\sum\limits_{|m| \le j}c_{|p|m}^{\frac{j}{2}} D_{jm, jm}^{(j, \gamma j)}
\end{equation}
, where $c_{|p|m}^{\frac{j}{2}}$ is $\phi(u)$'s Fourier transform:
\begin{equation}
\label{SU(2)-Fourier}
c_{|p|m}^{\frac{j}{2}} = (j+1)^{\frac{1}{2}}\int \limits_{SU(2)} \phi(u) \overline{D_{|p|m}^{\frac{j}{2}}(u)} \; du
\end{equation}
, where $p, j, m, n \in Z, \; \gamma \in C, \; u \in SU(2), \; g \in SL(2,C)$
Note that the parameter $j$ in $SL(2,C)$ matrix coefficients is an integer, while the parameter of the $SU(2)$ in $c_{|p|m}^{\frac{j}{2}}$ is half-integer.
\\[4ex]
\textbf{Theorem 1 -  SU(2)-Y-Map Existence Theorem: }\\[2ex]  
The sum $ \phi(u) \rightarrow \psi(g) = \quad \sum\limits_{j=|p|}^{\infty}\sum\limits_{|m| \le j} c^{\frac{j}{2}}_{|p|m} D^{(j, \gamma j)}_{jm, jm}(g) $ is convergent. \\[2ex]
\textbf{Proof:}\\[2ex]
By Paley-Wiener Theorem ($\cite{Ruhl}$ page 60, 91, see also $\cite{Gelfand}$) the Fourier transform $ c^{\frac{j}{2}}_{|p|m}$ satisfies the following asymptotic inequality:
$ \forall k \in N, k \ge 1$ or we can rewrite it as:
\begin{equation}
\label{asymptotic3}
|c^{\frac{j}{2}}_{|p|m}| \le \frac{C_k}{|j|^k}
\end{equation}
which means that the Fourier transform is a fast dropping function and decreases faster than any polynomial of power $k$, where $C_k$ is a constant depending on k only. Therefore
\begin{multline}
\sum\limits_{j=|p|}^{\infty}\sum\limits_{|m| \le j} c^{\frac{j}{2}}_{|p|m} D^{(j, \gamma j)}_{jm, jm}(g) \le  \sum\limits_{j=|p|}^{\infty}  \sum\limits_{|m| \le j} |c^{\frac{j}{2}}_{|p|m} D^{(j, \gamma j)}_{jm,  jm}(g)| \le \sum\limits_{j=|p|}^{\infty}  \sum\limits_{|m| \le j} |\frac{C_k}{|j|^k} D^{(j, \gamma j)}_{jm,  jm}(g)| \\
\le |C_k|\sum\limits_{j=|p|}^{\infty}  \sum\limits_{|m| \le j} |\frac{ D^{(j, \gamma j)}_{jm,  jm}(g)}{j^k}|
\end{multline}
and the last sum is convergent by Lemma1: $\sum\limits_{j=1}^{\infty}  \sum\limits_{|m| \le j} \frac{ D^{(j, \gamma j)}_{jm,  jm}(g)}{j^k}$ is absolute convergent and therefore also convergent for any $k \in N, k \ge 2$. See the proof of the Lemma 1 below. In Appendix B we also provide the graph from the numerical calculations by using MPMath Python program to demonstrate visually convergence stated in Lemma 1. \\
The limit is a function on $SL(2,C)$  since each $g \in SL(2,C)$ we map to the sum limit and the limit is unique by construction.\\[4ex]
$\square$\\[2ex]
SU(2)-Y-Map Existence Theorem establishes a map from the space of square integrable functions on $SU(2)$ to the space of  functions (not necessarily square integrable) on $SL(2,C)$.\\[2ex]
\textbf{Lemma 1:} $\quad$ For $k \ge 2$ the sum $\sum\limits_{j=1}^{\infty}  \sum\limits_{|m| \le j} \frac{ D^{(j, \tau j)}_{jm,  jm}(g)}{j^k}$ is absolute convergent and therefore convergent for all $g \in SL(2,C), \; \tau \in C, j, m \in Z$. \\[2ex]
\textbf{Proof:}\\[2ex]
According to the D'Alembert ratio test we need to prove:
\begin{equation}
\label{DAlembert2}
 \lim\limits_{j \rightarrow \infty}  \left|\frac{ \sum\limits_{|m| \le j+1}j^k D^{((j+1),\tau (j+1))}_{(j+1)m, \; (j+1)m}(g)}{ \sum\limits_{|m| \le j}  (j+1)^k D^{(j, \tau j)}_{jm,  jm}(g)}\right| < 1
\end{equation}
 Let us use the explicit expression for the matrix coefficients in ($\ref{DAlembert2}$). The first explicit expression of the principal series matrix coefficients $D^{(k', \rho)}_{jn, j'm}, k' \in Z, \rho \in C $ was obtained by Duc and Hieu in 1967  $\cite{DucHieu}$, formula (4.11):
\begin{multline}
\label{SL2CMatrix}
D^{(k', \rho)}_{jm, j'n}(g) =\frac{\delta_{mn}}{(j+j'+1)!}\\
{\left( (2j+1)(2j'+1)(j+m)!(j'+m)!(j-m)!(j'-m)!(j+k')!(j'+k')!(j-k')!(j'-k')! \right) }^{1/2}\\
\times \sum\limits_{d, d'} {(-1)}^{d + d'}\frac{(d+d'+m+k')!(j+j'-d-d'-m-k')!}{d!d'!(j-m-d)!(j'-m-d')!(k'+m+d)!(k' + m + d')!(j-k'-d)!(j'-k'-d')!}\\
\times {\epsilon}^{2(2d' + m + k' + 1 + \frac{i \rho}{2})}{}_2F_1( j' + 1 + \frac{i \rho}{2}, d+d'+m+k'+1; j+j'+2; 1-{\epsilon}^4)
\end{multline}
,where ${}_2F_1(\alpha, \beta; \gamma; z)$ - is a hypergeometric function, $d$ and $d'$ are integers that do not make each factor under the factorial to become a negative number and  $\epsilon$ is a real number obtained from the $g \in SL(2,C)$ decomposition:
\begin{equation}
\label{gdecomposition}
g=u_1 b u_2 
\end{equation}
,where $u_1$ and $u_2$ are unitary matrices, while the matrix $b$ =
$
\begin{pmatrix}
\label{matrix}
{\epsilon}^{-1}  & 0 \\
0 & \epsilon  \\
\end{pmatrix}
$, $\epsilon \in R$\\[2ex]
As one can see all $D^{(k', \rho)}_{jm, j'n}(g)$ are zero for $m \neq n$ due to the presence of the Kronecker delta in ($\ref{SL2CMatrix}$). Therefore we can omit all zero terms in the sums and leave only the terms with $m=n$. That's why we wrote our sum only over $j$ and $m$:  
\begin{equation}
\label{Psig2}
 \sum\limits_{j=1}^{\infty}\sum\limits_{|m| \le j} \frac{D^{(j, \tau j)}_{jm,  jm}(g)}{j^k}
\end{equation}
The matrix coefficients in our sum  $D^{(j, \tau j)}_{jm,  jm}(g)$ have much simpler form than the general form ($\ref{SL2CMatrix}$).
We rewrite the unitary matrix coefficients $D^{(k', \rho)}_{jm, j'n}(g)$ in ($\ref{SL2CMatrix}$) for:
$k' = j, \; \rho = \tau j,  \; j' = j, \; m=n$. Also since $d$ and $d'$ are so that factorial expressions are non-negative, one can see from  ($\ref{SL2CMatrix}$) that if $k' = j$, which is our case, then  $j-k' - d \ge 0$ implies $j-j - d \ge 0$, so $d \le 0$, but at the same time $d!$ implies $d \ge 0$ so it follows that $d=0$. The same is true for $d' = 0$ and the sums over $d$ and $d'$ in ($\ref{SL2CMatrix}$) disappear:
\begin{multline}
\label{SL2CMatrix1}
D^{(j, \tau j)}_{jm, jm}(g) =\frac{1}{(2j+1)!} (2j)!(2j+1)(j+m)!(j-m)!\times \frac{(j+m)!(j-m)!}{(j-m)!(j-m)!(j+m)!(j + m)!}\\
\times {\epsilon}^{2(m + j + 1 + \frac{i \tau j}{2})}{}_2F_1( j + 1 + \frac{i \tau j}{2}, m+j+1; 2j+2; 1-{\epsilon}^4)
\end{multline}
All coefficients cancel as one can see and we obtain:
\begin{equation}
\label{SL2CMatrix2}
D^{(j, \tau j)}_{jm, jm}(g) = {\epsilon}^{2(m + j + 1 + \frac{i \tau j}{2})} {}_2F_1( j + 1 + \frac{i \tau j}{2}, m+j+1; 2j+2; 1-{\epsilon}^4)
\end{equation}
The sum ($\ref{Psig2}$) becomes:
\begin{equation}
\label{NewSum}
\sum\limits_{j=1}^{\infty}\sum\limits_{|m| \le j}\frac{1}{j^k}  D^{(j, \tau j)}_{jm, jm}(g) = \sum\limits_{j=1}^{\infty}\sum\limits_{|m| \le j} \frac{1}{j^k} {\epsilon}^{2(m + j + 1 + \frac{i \tau j}{2})} {}_2F_1( j + 1 + \frac{i \tau j}{2}, m+j+1; 2j+2; 1-{\epsilon}^4) 
\end{equation}
We now consider the following two sums: first for $ 0 \le m \le j$  and the second for  $  -j \le m < 0 $ and by bounding them from above we will prove their convergence. The convergence of the original sum will then follow. 
\begin{multline}
\label{Bounding}
\sum\limits_{j=1}^{\infty}\frac{1}{j^k}\left | \sum\limits_{|m| \le j} D^{(j, \tau j)}_{jm, jm}(g) \right | \le \\
\sum\limits_{j=1}^{\infty} \frac{1}{j^k} \left | \sum\limits_{m = 0}^{m=j}  {\epsilon}^{2(m + j + 1 + \frac{i \tau j}{2})} {}_2F_1( j + 1 + \frac{i \tau j}{2}, m+j+1; 2j+2; 1-{\epsilon}^4)  \right | + \\
\sum\limits_{j=1}^{\infty} \frac{1}{j^k} \left | \sum\limits_{m = -j}^{m < 0}  {\epsilon}^{2(m + j + 1 + \frac{i \tau j}{2})} {}_2F_1( j + 1 + \frac{i \tau j}{2}, m+j+1; 2j+2; 1-{\epsilon}^4)  \right | \le \\
\sum\limits_{j=1}^{\infty}\frac{1}{j^k}\sum\limits_{m = 0}^{m=j} \left | {\epsilon}^{2(m + j + 1 + \frac{i \tau j}{2})} \right | \left |{}_2F_1( j + 1 + \frac{i \tau j}{2}, m+j+1; 2j+2; 1-{\epsilon}^4)  \right | + \\
\sum\limits_{j=1}^{\infty}\frac{1}{j^k}\sum\limits_{m = -j}^{m < 0} \left | {\epsilon}^{2(m + j + 1 + \frac{i \tau j}{2})} \right | \left |{}_2F_1( j + 1 + \frac{i \tau j}{2}, m+j+1; 2j+2; 1-{\epsilon}^4)  \right | \le \\
\sum\limits_{j=1}^{\infty} \frac{1}{j^k}  \left | (j+1) {\epsilon}^{2(j + j + 1 + \frac{i \tau j}{2})} \right | \left |{}_2F_1( j + 1 + \frac{i \tau j}{2}, j+j+1; 2j+2; 1-{\epsilon}^4)  \right | + \\
\sum\limits_{j=1}^{\infty} \frac{1}{j^k}  \left | j {\epsilon}^{2(0 + j + 1 + \frac{i \tau j}{2})} \right | \left |{}_2F_1( j + 1 + \frac{i \tau j}{2}, 0+j+1; 2j+2; 1-{\epsilon}^4)  \right |
\end{multline}
We pass to the last inequality above by putting $m = j$ in the first sum and $m = 0$ in the second and remembering  the hypergeometric function is monotonic with respect to its second argument:
\begin{equation}
\label{Hypergeometric}
{}_2F_1(a, b; c; z) = \sum\limits_{n=0}^{\infty} \frac{{(a)}_n {(b)}_n z^n}{{(c)}_n n!}\\
\end{equation}
,where 
\begin{equation}
{(q)}_n = 1, \mbox{when} \; n =0, {(q)}_n = q(q+1) \mbox{...} (q + n -1),   n > 0
\end{equation}
The hypergeometric function is originally defined for $|z| <1 $, but is analytically continued to all values of $z$ as was shown in $\cite{Watson}$.\\[2ex]
In our case of   ${}_2F_1( j + 1 + \frac{i \tau j}{2}, m+j+1; 2j+2; 1-{\epsilon}^4)$,  \;
the parameter $ b = m + j + 1 $ is always positive and the absolute value of the function is increasing when m is increasing. That is why in the last inequality of ($\ref{Bounding}$) we put $m = j$ to bound the sum from above when $ m \ge 0$ and by $m = 0$ in the second sum, when $m < 0$.\\[2ex]
At this point we are going to use the D'Alembert ratio convergence test and the asymptotic of the hypergeometric function to prove that the two bounding from above sums are convergent and that will prove that the original sum is convergent.
We will need to consider three cases: $|\epsilon| < 1$, $|\epsilon| > 1$, $\epsilon = 1$ \\[2ex]
The hypergeometric function ${}_2F_1(\alpha, \beta; \gamma; y)$ asymptotic, when all three parameters go to infinity, was investigated and derived by G.N Watson (1918) and can be found in Bateman's book $\cite{Bateman}$ volume 1 page 77:
\begin{multline}
\label{Asymptotic}
{\left(\frac{z}{2} -\frac{1}{2}\right)}^{-a -\lambda}{}_2F_1(a +\lambda, a - c + 1 + \lambda; a-b+1+2\lambda; 2{(1-z)}^{-1}) =\\ \frac{2^{a+b}\Gamma(a-b+1+2\lambda)\Gamma(1/2){\lambda}^{-1/2}}{\Gamma(a-c+1+\lambda)\Gamma(c-b+\lambda)} e^{-(a+\lambda)\xi}\times {(1-e^{-\xi})}^{-c+1/2} \times {(1+e^{-\xi})}^{c-a-b-1/2}[1+ O({\lambda}^{-1})]
\end{multline}
,where $\xi$ is defined as following: $e^{\pm \xi} = z \pm \sqrt{z^2-1}$. The minus sign corresponds to $Im(z) \le 0$, the plus sign to $Im(z) > 0$. This asymptotic also works in the limit case of z being real, which is our case of $1-{\epsilon}^4$ (for details see Watson's original 1918 paper $\cite{Watson}$) \\[2ex]
By comparing ($\ref{SL2CMatrix2}$) and $(\ref{Asymptotic})$  we see that the  hypergeometric function arguments $\lambda, a, b, c$  in our case take the following values:
\begin{equation}
\label{Parameters}
\lambda = j, \; a = 1 + \frac{i \tau j}{2}, \; b = \frac{i \tau j}{2}, \; c = 1+\frac{i \tau j}{2} - m, \; z = \frac{{\epsilon}^4 + 1}{{\epsilon}^4 -1}, \;  e^{\mp \xi} = \frac{{\epsilon}^2 \mp 1}{{\epsilon}^2 \pm 1}
\end{equation}
Indeed by substituting them into l.h.s of the ($\ref{Asymptotic}$) we get ${}_2F_1$ exactly as in $(\ref{SL2CMatrix2})$:\\[2ex]
${}_2F_1( j + 1 + \frac{i \tau j}{2}, m+j+1; 2j+2; 1-{\epsilon}^4)$\\[2ex]
Let us rewrite ($\ref{Asymptotic}$) then in terms of  $(j, m, \tau)$ and we obtain:
\begin{multline}
\label{Asymptotic1}
{}_2F_1( j + 1 + \frac{i \tau j}{2}, m+j+1; 2j+2; 1-{\epsilon}^4) = \\
\frac{1}{{({\epsilon}^4-1)}^{1+j+\frac{i \tau j}{2}}}
\frac{2^{(1+ i \tau j)}\Gamma(2 + 2j)\Gamma(\frac{1}{2})j^{-1/2}}{\Gamma(m+1+j) \Gamma(1-m+j)}\times \\
e^{-(1+\frac{i \tau j}{2} + j)\xi} \times  (1-e^{-\xi})^{(-\frac{1}{2}-\frac{i \tau j}{2}+m)} \times  (1+e^{-\xi})^{(-m -\frac{i \tau j}{2}-\frac{1}{2})}\left[ 1 + O(\frac{1}{j})\right]
\end{multline}
or by expressing $e^{-\xi}$ in terms of $\epsilon$ by using ($\ref{Parameters}$) we obtain the following expression:
\begin{multline}
\label{Asymptotic2}
{}_2F_1( j + 1 + \frac{i \tau j}{2}, m+j+1; 2j+2; 1-{\epsilon}^4) = \\
\frac{1}{{({\epsilon}^4-1)}^{1+j+\frac{i \tau j}{2}}}
\frac{2^{(1+ i \tau j)}\Gamma(2 + 2j)\Gamma(\frac{1}{2})j^{-1/2}}{\Gamma(m+1+j) \Gamma(1-m+j)}\times \\
\left ( {\frac{{\epsilon}^2 -1}{{\epsilon}^2 +1}}\right )^{(1+\frac{i \tau j}{2} + j)} \times  \left ( {\frac{2}{{\epsilon}^2 +1}} \right ) ^{(-\frac{1}{2}-\frac{i \tau j}{2}+m)}\times  \left ( {\frac{2{\epsilon}^2}{{\epsilon}^2 + 1}} \right ) ^{(-m -\frac{i \tau j}{2}-\frac{1}{2})}\left[ 1 + O(\frac{1}{j})\right]
\end{multline}
We rewrite this expression by denoting the right hand side before $\left[ 1 + O(\frac{1}{j})\right]$ as ${}_2A_1(j, m, \tau, \epsilon)$.
\begin{equation}
\label{Asymptotic22}
{}_2F_1( j + 1 + \frac{i \tau j}{2}, m+j+1; 2j+2; 1-{\epsilon}^4) ={}_2A_1(j, m, \tau, \epsilon) \left[ 1 + O(\frac{1}{j})\right]
\end{equation}
We are going to use this expression in the D'Alembert ratio test to prove the convergence of the bounding sums in ($\ref{Bounding}$). The first sum corresponds to $m = j$

\begin{multline}
\label{FirstSum}
\sum\limits_{j=1}^{\infty}\frac{1}{j^k} \left | (j+1) {\epsilon}^{2(j + j + 1 + \frac{i \tau j}{2})} \right | \left |{}_2F_1( j + 1 + \frac{i \tau j}{2}, j+j+1; 2j+2; 1-{\epsilon}^4)  \right | 
\end{multline}

while the second to $m=0$:

\begin{multline}
\label{SecondSum}
\sum\limits_{j=1}^{\infty} \frac{1}{j^k}  \left | j {\epsilon}^{2(0 + j + 1 + \frac{i \tau j}{2})} \right | \left |{}_2F_1( j + 1 + \frac{i \tau j}{2}, 0+j+1; 2j+2; 1-{\epsilon}^4)  \right |
\end{multline}
Before we begin, we can see right away that $ O(\frac{1}{j})$ in ($\ref{Asymptotic22}$) creates a problem for applying a ratio test. However it can be easily fixed. We substitute ($\ref{Asymptotic22}$)  into ($\ref{FirstSum}$)  and obtain:
\begin{multline}
\label{FirstSum1}
\sum\limits_{j=1}^{\infty}\frac{1}{j^k} \left | (j+1) {\epsilon}^{2(j + j + 1 + \frac{i \tau j}{2})} \right | \left|{}_2A_1(j, m, \tau, \epsilon)  \left[ 1 + O(\frac{1}{j})\right] \right| \le \\
\sum\limits_{j=1}^{\infty}\left | (j+1) {\epsilon}^{2(j + j + 1 + \frac{i \tau j}{2})} \right | \left|{}_2A_1(j, m, \tau, \epsilon) \right|\frac{1}{j^k}  + \\
\sum\limits_{j=1}^{\infty}\left | (j+1) {\epsilon}^{2(j + j + 1 + \frac{i \tau j}{2})} \right | \left|{}_2A_1(j, m, \tau, \epsilon) \right | \left[ O(\frac{1}{j^{k+1}})\right]  
\end{multline}
If we prove that $\sum\limits_{j=1}^{\infty}\left | (j+1) {\epsilon}^{2(j + j + 1 + \frac{i \tau j}{2})} \right | \left|{}_2A_1(j, m, \tau, \epsilon) \right|$ is convergent, then the first sum on the right hand side will be convergent for $k \ge 1$ by Abel's theorem, stating that if $\sum a_n$ is convergent sequence, and $b_n$ is monotonic bounded , then $\sum a_nb_n$ is convergent, while the second sum on the right will be convergent as $ O(\frac{1}{j^{k+1}})$ is convergent for $k \ge 1$, so it will be bounded by the product of two converging series: .
\begin{multline}
\sum\limits_{j=1}^{\infty}\left | (j+1) {\epsilon}^{2(j + j + 1 + \frac{i \tau j}{2})} \right | \left|{}_2A_1(j, m, \tau, \epsilon) \right | \left[ O(\frac{1}{j^{k+1}})\right]  \le \\
 \sum\limits_{j=1}^{\infty}\left | (j+1) {\epsilon}^{2(j + j + 1 + \frac{i \tau j}{2})} \right | \left|{}_2A_1(j, m, \tau, \epsilon) \right |  \times  \sum\limits_{j=1}^{\infty}\left | \left[ O(\frac{1}{j^{k+1}})\right] \right|
\end{multline} 
The same logic applies to the second bounding sum ($\ref{SecondSum}$). \\
Therefore, in order to prove the bounding sums  ($\ref{FirstSum}$), ($\ref{SecondSum}$) convergence it is enough to prove the convergence of:
\begin{equation}
\label{FirstSum3}
\sum\limits_{j=1}^{\infty}\left | (j+1) {\epsilon}^{2(j + j + 1 + \frac{i \tau j}{2})} \right | \left|{}_2A_1(j, m, \tau, \epsilon) \right|
\end{equation}
and
\begin{equation}
\label{SecondSum3}
\sum\limits_{j=1}^{\infty}  \left | j {\epsilon}^{2(0 + j + 1 + \frac{i \tau j}{2})} \right | \left |{}_2A_1(j,m, \tau, \epsilon )  \right |
\end{equation}
, where we remind that by definition from ($\ref{Asymptotic2}$)
\begin{multline}
\label{ADefinition}
{}_2A_1(j,m, \tau, \epsilon ) = {}_2A_1( j + 1 + \frac{i \tau j}{2}, m+j+1; 2j+2; 1-{\epsilon}^4) = \\
 \frac{1}{{({\epsilon}^4-1)}^{1+j+\frac{i \tau j}{2}}}
\frac{2^{(1+ i \tau j)}\Gamma(2 + 2j)\Gamma(\frac{1}{2})j^{-1/2}}{\Gamma(m+1+j) \Gamma(1-m+j)}\times \\
\left ( {\frac{{\epsilon}^2 -1}{{\epsilon}^2 +1}}\right )^{(1+\frac{i \tau j}{2} + j)} \times  \left ( {\frac{2}{{\epsilon}^2 +1}} \right ) ^{(-\frac{1}{2}-\frac{i \tau j}{2}+m)}\times  \left ( {\frac{2{\epsilon}^2}{{\epsilon}^2 + 1}} \right ) ^{(-m -\frac{i \tau j}{2}-\frac{1}{2})}
\end{multline}
We used two notations for the same function ${}_2A_1$ above. The second one is for showing the arguments explicitly, which is more convenient, when we begin using the ratio tests below.\\
By proving the sums convergence we would need to consider two cases of $|\epsilon| > 1$ and $|\epsilon| < 1$ for each sum separately, i.e. four cases all together. The simple fifth case $\epsilon = 1$ is considered at the end.\\[2ex]
\textbf{Case 1:}  First sum, $m = j$, $\tau \in C, \tau = \eta + i \omega$, $|\epsilon| > 1$\\[2ex]
\begin{equation}
\sum\limits_{j=1}^{\infty} \left | (j+1) {\epsilon}^{2(2j + 1 + \frac{i \tau j}{2})} \right | \left |{}_2A_1( j + 1 + \frac{i \tau j}{2}, 2j+1; 2j+2; 1-{\epsilon}^4)  \right |
\end{equation}
The D'Alembert ratio test is as follows:
\begin{multline}
\lim\limits_{j \rightarrow \infty} \left | \frac{(j+2) {\epsilon}^{2(2(j+1) + 1 + \frac{i \tau (j+1)}{2})}}{(j+1) {\epsilon}^{2(2j+ 1 + \frac{i \tau j}{2})}} \right | \left |\frac{{}_2A_1( j + 2 + \frac{i \tau (j+1)}{2}, 2(j+1)+1; 2(j+1)+2; 1-{\epsilon}^4)}{{}_2A_1( j + 1 + \frac{i \tau j}{2}, 2j+1; 2j+2; 1-{\epsilon}^4)}  \right | = \\
{\epsilon}^4 {\epsilon}^{-\omega}\times \lim\limits_{j \rightarrow \infty} \left |\frac{{}_2A_1( j + 2 + \frac{i \tau (j+1)}{2}, 2(j+1)+1; 2(j+1)+2; 1-{\epsilon}^4)}{{}_2A_1( j + 1 + \frac{i \tau j}{2}, 2j+1; 2j+2; 1-{\epsilon}^4)} \right |
\end{multline}
by using  ($\ref{ADefinition}$) for  $m = j$ we obtain:
\begin{multline}
 \lim\limits_{j \rightarrow \infty} {\epsilon}^4 {\epsilon}^{-\omega}\times \left |\frac{{}_2A_1( j + 2 + \frac{i \tau (j+1)}{2}, 2(j+1)+1; 2(j+1)+2; 1-{\epsilon}^4)}{{}_2A_1( j + 1 + \frac{i \tau j}{2}, 2j+1; 2j+2; 1-{\epsilon}^4)} \right | = \\
 \lim\limits_{j \rightarrow \infty} {\epsilon}^4 {\epsilon}^{-\omega}\times \left |  \frac{{({\epsilon}^4-1)}^{1+j+\frac{i \tau j}{2}}}{{({\epsilon}^4-1)}^{1+j+1+\frac{i \tau (j+1)}{2}}}\frac{2^{(1+i\tau (j+1))}}{2^{(1+i\tau j)}}\frac{\Gamma(2 + 2(j+1))\Gamma(\frac{1}{2}){(j+1)}^{-1/2}\Gamma(2j+1)\Gamma(1)}{\Gamma(2+2j)\Gamma(\frac{1}{2})j^{-1/2}\Gamma(2(j+1)+1) \Gamma(1)} \right | \times  \\
\left | \left ( {\frac{{\epsilon}^2 -1}{{\epsilon}^2 +1}}\right )^{(1+\frac{i \tau (j+1)}{2} + (j+1))-(1+\frac{i \tau j}{2} + j)} \right |  \times \left | \left ( {\frac{2}{{\epsilon}^2 +1}} \right ) ^{(-\frac{1}{2}-\frac{i \tau (j+1)}{2}+(j+1))-(-\frac{1}{2}-\frac{i \tau j}{2}+j)} \right |  \\
\times  \left | \left ( {\frac{2{\epsilon}^2}{{\epsilon}^2 + 1}} \right ) ^{(-j -1 -\frac{i \tau (j+1)}{2}-\frac{1}{2})-(-j -\frac{i \tau j}{2}-\frac{1}{2})} \right | = \\
  \lim\limits_{j \rightarrow \infty}  \left | \frac{{\epsilon}^4 {\epsilon}^{-\omega} 2^{-\omega}}{({\epsilon}^4-1)^{(1-\frac{\omega}{2})}}   \frac{(2j+3)(2j+2)\Gamma(2j+2)\Gamma(2j+1)}{(2j+2)(2j+1)\Gamma(2j+2)\Gamma(2j+1)}    \frac{{({\epsilon}^2 -1)}^{(1-\frac{\omega}{2})}}{{({\epsilon}^2 +1)}^{(1-\frac{\omega}{2})}}\frac{2^{(1+\frac{\omega}{2})}}{{({\epsilon}^2+1)}^{(1+\frac{\omega}{2})}} \frac{{({\epsilon}^2 + 1)}^{(1-\frac{\omega}{2})}}{{(2{\epsilon}^2)}^{(1-\frac{\omega}{2})}} \right | = \\
\frac{{\epsilon}^2}{{({\epsilon}^2 + 1)}^2} < 1,  \;  \forall |\epsilon| > 1 
\end{multline}
 We used the fact that the absolute value of the positive real number in the pure imaginary power is 1 and the property of the $\Gamma$ function: $\Gamma(z+1) = z \Gamma(z)$. By this property all $\Gamma$ above cancel. We also remind that $\omega$ in the formula above comes from $\tau = \eta + i \omega$. \\[2ex]

\textbf{Case 2:}  First sum $m = j$, $\tau \in C, \tau = \eta + i \omega$, $|\epsilon| < 1$\\[2ex]
\begin{equation}
\sum\limits_{j=1}^{\infty} \left | (j+1) {\epsilon}^{2(2j + 1 + \frac{i \tau j}{2})} \right | \left |{}_2A_1( j + 1 + \frac{i \tau j}{2}, 2j+1; 2j+2; 1-{\epsilon}^4)  \right |
\end{equation}
The D'Alembert ratio test provides the expression very similar to the Case 1 with one difference. In this case of $|\epsilon| < 1$
we write the following expressions in the form:
\begin{equation}
 {\epsilon}^4 - 1 = (1 - {\epsilon}^4) e^{\pm i \pi}
\end{equation} 
\begin{equation}
{\epsilon}^2 - 1 = (1 - {\epsilon}^2) e^{\pm i \pi}
\end{equation}
\begin{multline}
 \lim\limits_{j \rightarrow \infty} {\epsilon}^4 {\epsilon}^{-\omega} \times \left |\frac{{}_2A_1( j + 2 + \frac{i \tau (j+1)}{2}, 2(j+1)+1; 2(j+1)+2; 1-{\epsilon}^4)}{{}_2A_1( j + 1 + \frac{i \tau j}{2}, 2j+1; 2j+2; 1-{\epsilon}^4)} \right | = \\
 \lim\limits_{j \rightarrow \infty} {\epsilon}^4 {\epsilon}^{-\omega}\times \left |  \frac{{(({1-\epsilon}^4) e^{\pm i \pi})}^{1+j+\frac{i \tau j}{2}}}{{(({1-\epsilon}^4) e^{\pm i \pi}})^{1+j+1+\frac{i \tau (j+1)}{2}}}\frac{2^{(1+i\tau (j+1))}}{2^{(1+i\tau j)}} \frac{\Gamma(2 + 2(j+1))\Gamma(\frac{1}{2}){(j+1)}^{-1/2}\Gamma(2j+1)\Gamma(1)}{\Gamma(2+2j)\Gamma(\frac{1}{2})j^{-1/2}\Gamma(2(j+1)+1) \Gamma(1)} \right | \times  \\
\left | \left ( {\frac{(1-{\epsilon}^2) e^{\pm i \pi}}{{\epsilon}^2 +1}}\right )^{(1+\frac{i \tau (j+1)}{2} + (j+1))-(1+\frac{i \tau j}{2} + j)} \right | \times  \left | \left ( {\frac{2}{{\epsilon}^2 +1}} \right ) ^{(-\frac{1}{2}-\frac{i \tau (j+1)}{2}+(j+1))-(-\frac{1}{2}-\frac{i \tau j}{2}+j)} \right | \times \\ 
\left | \left ( {\frac{2{\epsilon}^2}{{\epsilon}^2 + 1}} \right ) ^{(-j -1 -\frac{i \tau (j+1)}{2}-\frac{1}{2})-(-j -\frac{i \tau j}{2}-\frac{1}{2})} \right | = \\
	  \lim\limits_{j \rightarrow \infty}  \left | \frac{{\epsilon}^4 {\epsilon}^{-\omega}2^{-\omega}}{(1-{\epsilon}^4)^{(1-\frac{\omega}{2})} e^{\mp \frac{\pi \eta}{2}}}  \frac{(2j+3)(2j+2)\Gamma(2j+2)\Gamma(2j+1)}{(2j+2)(2j+1)\Gamma(2j+2)\Gamma(2j+1)}    \frac{(1-{\epsilon}^2)^{(1-\frac{\omega}{2})}e^{\mp \frac{\pi \eta}{2}}}{({\epsilon}^2 +1)^{(1-\frac{\omega}{2})}}\frac{2^{(1+\frac{\omega}{2})}}{({\epsilon}^2+1)^{(1+\frac{\omega}{2})}} \frac{({\epsilon}^2 + 1)^{(1-\frac{\omega}{2})}}{(2{\epsilon}^2)^{(1-\frac{\omega}{2})}} \right | = \\ 
\frac{{\epsilon}^2}{{({\epsilon}^2 + 1)}^2} < 1, \; \forall |\epsilon| < 1
\end{multline}
\textbf{Case 3:}  Second sum,  $m = 0$, $\tau \in C, \tau = \eta + i \omega $, $|\epsilon| > 1$\\[2ex]
\begin{equation}
\sum\limits_{j=1}^{\infty} \left | j {\epsilon}^{2(j + 1 + \frac{i \tau j}{2})} \right | \left |{}_2A_1( j + 1 + \frac{i \tau j}{2}, j+1; 2j+2; 1-{\epsilon}^4)  \right |
\end{equation}
D'Alembert ratio test is as follows:
\begin{multline}
\lim\limits_{j \rightarrow \infty} \left | \frac{(j+1) {\epsilon}^{2(j+1+ 1 + \frac{i \tau (j+1)}{2})}}{j {\epsilon}^{2(j + 1 + \frac{i \tau j}{2})}} \right | \left |\frac{{}_2A_1( j + 2+ \frac{i \tau (j+1)}{2}, j+2; 2(j+1)+2; 1-{\epsilon}^4)}{{}_2A_1( j + 1 + \frac{i \tau j}{2}, j+1; 2j+2; 1-{\epsilon}^4)}  \right | = \\
{\epsilon}^2 {\epsilon}^{-\omega}\times \lim\limits_{j \rightarrow \infty} \left |\frac{{}_2A_1( j + 2+ \frac{i \tau (j+1)}{2}, j+2; 2(j+1)+2; 1-{\epsilon}^4)}{{}_2A_1( j + 1 + \frac{i \tau j}{2}, j+1; 2j+2; 1-{\epsilon}^4)}  \right |
\end{multline}
We use ($\ref{ADefinition}$) for $m = 0$ 
\begin{multline}
 \lim\limits_{j \rightarrow \infty}  {\epsilon}^2 {\epsilon}^{-\omega}\left |\frac{{}_2A_1( j + 2+ \frac{i \tau (j+1)}{2}, j+2; 2(j+1)+2; 1-{\epsilon}^4)}{{}_2A_1( j + 1 + \frac{i \tau j}{2}, j+1; 2j+2; 1-{\epsilon}^4)}  \right | = \\
 \lim\limits_{j \rightarrow \infty} {\epsilon}^2 {\epsilon}^{-\omega} \times \left |  \frac{{({\epsilon}^4-1)}^{1+j+\frac{i \tau j}{2}}}{{({\epsilon}^4-1)}^{1+j+1+\frac{i \tau (j+1)}{2}}}\frac{2^{(1+i\tau (j+1))}}{2^{(1+i\tau j)}} \frac{\Gamma(2 + 2(j+1))\Gamma(\frac{1}{2}){(j+1)}^{-1/2}\Gamma(j+1)\Gamma(j+1)}{\Gamma(2+2j)\Gamma(\frac{1}{2})j^{-1/2}\Gamma(j+2) \Gamma(j+2)} \right | \times  \\
\left | \left ( {\frac{{\epsilon}^2 -1}{{\epsilon}^2 +1}}\right )^{(1+\frac{i \tau (j+1)}{2} + (j+1))-(1+\frac{i \tau j}{2} + j)} \right | \times \left | \left ( {\frac{2}{{\epsilon}^2 +1}} \right ) ^{(-\frac{1}{2}-\frac{i \tau (j+1)}{2})-(-\frac{1}{2}-\frac{i \tau j}{2})} \right | \\
 \times \left | \left ( {\frac{2{\epsilon}^2}{{\epsilon}^2 + 1}} \right ) ^{( -\frac{i \tau (j+1)}{2}-\frac{1}{2})-( -\frac{i \tau j}{2}-\frac{1}{2})} \right | = \\
  \lim\limits_{j \rightarrow \infty}  \left | \frac{{\epsilon}^2{\epsilon}^{-\omega}2^{-\omega}}{({\epsilon}^4-1)^{(1-\frac{\omega}{2})}}  \frac{(2j+3)(2j+2)\Gamma(2j+2)\Gamma(j+1)\Gamma(j+1)}{(j+1)(j+1)\Gamma(2j+2)\Gamma(j+1)\Gamma(j+1)}    \frac{({\epsilon}^2 -1)^{(1-\frac{\omega}{2})}}{({\epsilon}^2 +1)^{(1-\frac{\omega}{2})}} \frac{2^{\frac{\omega}{2}}}{({\epsilon}^2 + 1)^{\frac{\omega}{2}}} \frac{({2{\epsilon}^2)}^{\frac{\omega}{2}}}{({\epsilon}^2 + 1)^{\frac{\omega}{2}}} \right | = \\
\frac{4{\epsilon}^2}{{({\epsilon}^2 + 1)}^2} < 1, \;  \forall |\epsilon| > 1
\end{multline}

\textbf{Case 4:}  Second sum, $m = 0$, $\tau \in C, \tau = \eta + i \omega,  |\epsilon| < 1$\\[2ex]
\begin{equation}
\sum\limits_{j=1}^{\infty} \left | j {\epsilon}^{2(j + 1 + \frac{i \tau j}{2})} \right | \left | {}_2A_1( j + 1 + \frac{i \tau j}{2}, j+1; 2j+2; 1-{\epsilon}^4)\right |
\end{equation}
This case is similar to Case 3. For $|\epsilon| < 1$ we need to write again the following two expressions in the form:
\begin{equation}
 {\epsilon}^4 - 1 = (1 - {\epsilon}^4) e^{\pm i \pi}
\end{equation} 
\begin{equation}
{\epsilon}^2 - 1 = (1 - {\epsilon}^2) e^{\pm i \pi}
\end{equation}

The D'Alembert ratio test is as follows:
\begin{multline}
\lim\limits_{j \rightarrow \infty} \left | \frac{(j+1) {\epsilon}^{2(j+1+ 1 + \frac{i \tau (j+1)}{2})}}{j {\epsilon}^{2(j + 1 + \frac{i \tau j}{2})}} \right | \left |\frac{{}_2A_1( j + 2+ \frac{i \tau (j+1)}{2}, j+2; 2(j+1)+2; 1-{\epsilon}^4)}{{}_2A_1( j + 1 + \frac{i \tau j}{2}, j+1; 2j+2; 1-{\epsilon}^4)}  \right | = \\
{\epsilon}^2 {\epsilon}^{-\omega} \times \lim\limits_{j \rightarrow \infty} \left |\frac{{}_2A_1( j + 2+ \frac{i \tau (j+1)}{2}, j+2; 2(j+1)+2; 1-{\epsilon}^4)}{{}_2A_1( j + 1 + \frac{i \tau j}{2}, j+1; 2j+2; 1-{\epsilon}^4)}  \right |
\end{multline}
By using  ($\ref{ADefinition}$) for $m=0$ and $|\epsilon| < 1$
\begin{multline}
 \lim\limits_{j \rightarrow \infty}  {\epsilon}^2 {\epsilon}^{-\omega} \left |\frac{{}_2A_1( j + 2+ \frac{i \tau (j+1)}{2}, j+2; 2(j+1)+2; 1-{\epsilon}^4)}{{}_2A_1( j + 1 + \frac{i \tau j}{2}, j+1; 2j+2; 1-{\epsilon}^4)}  \right | = \\
 \lim\limits_{j \rightarrow \infty} {\epsilon}^2 {\epsilon}^{-\omega} \times \left |  \frac{({(1-{\epsilon}^4)e^{\pm i \pi})}^{1+j+\frac{i \tau j}{2}}}{{(({1-\epsilon}^4) e ^{\pm i \pi})}^{1+j+1+\frac{i \tau (j+1)}{2}}}\frac{2^{(1+i\tau (j+1))}}{2^{(1+i\tau j)}} \frac{\Gamma(2 + 2(j+1))\Gamma(\frac{1}{2}){(j+1)}^{-1/2}\Gamma(j+1)\Gamma(j+1)}{\Gamma(2+2j)\Gamma(\frac{1}{2})j^{-1/2}\Gamma(j+2) \Gamma(j+2)} \right | \times  \\
\left | \left ( {\frac{(1-{\epsilon}^2) e^{\pm i \pi} }{{\epsilon}^2 +1}}\right )^{(1+\frac{i \tau (j+1)}{2} + (j+1))-(1+\frac{i \tau j}{2} + j)} \right | \times  \left | \left ( {\frac{2}{{\epsilon}^2 +1}} \right ) ^{(-\frac{1}{2}-\frac{i \tau (j+1)}{2})-(-\frac{1}{2}-\frac{i \tau j}{2})} \right | \\
\times \left | \left ( {\frac{2{\epsilon}^2}{{\epsilon}^2 + 1}} \right ) ^{( -\frac{i \tau (j+1)}{2}-\frac{1}{2})-( -\frac{i \tau j}{2}-\frac{1}{2})} \right | = \\
  \lim\limits_{j \rightarrow \infty}  \left | \frac{{\epsilon}^2 {\epsilon}^{-\omega} 2^{-{\omega}}}{(1-{\epsilon}^4)^{(1-\frac{\omega}{2})}e^{\mp \frac{\pi \eta}{2}}}  \frac{(2j+3)(2j+2)\Gamma(2j+2)\Gamma(j+1)\Gamma(j+1)}{(j+1)(j+1)\Gamma(2j+2)\Gamma(j+1)\Gamma(j+1)}    \frac{(1-{\epsilon}^2 )^{(1-\frac{\omega}{2})} e^{\mp \frac{\pi \eta}{2}}}{({\epsilon}^2 +1)^{(1-\frac{\omega}{2})}}  \frac{2^{\frac{\omega}{2}}}{({\epsilon}^2 + 1)^{\frac{\omega}{2}}} \frac{({2{\epsilon}^2)}^{\frac{\omega}{2}}}{({\epsilon}^2 + 1)^{\frac{\omega}{2}}} \right | =\\ \frac{4{\epsilon}^2}{{({\epsilon}^2 + 1)}^2} < 1, \quad \forall |\epsilon| < 1 
\end{multline}
\textbf{Case 5:} $\epsilon = 1$\\
In the remaining case $\epsilon = 1$ we get $ {}_2F_1(a, b; c; 0) = 1 $ and $D^{(j, \tau j)}_{jm,  jm}(g) =1$ therefore 
$\sum\limits_{j=1}^{\infty}\sum\limits_{|m| \le j} \frac{1}{j^k}D^{(j, \tau j)}_{jm,  jm}(g) =\sum\limits_{j=1}^{\infty} \frac{2j+1}{j^k} $ is convergent for $k \ge 2$, as it is a Riemann zeta function.\\[2ex]
Thus, we have proved by ratio test that the limiting sums ($\ref{FirstSum3}$) and ($\ref{SecondSum3}$) are convergent, and hence by ($\ref{FirstSum1}$) trick, the limiting sums ($\ref{FirstSum}$) and ($\ref{SecondSum}$) are convergent, and therefore by ($\ref{Bounding}$) $\sum\limits_{j=1}^{\infty}  \sum\limits_{|m| \le j} \frac{ D^{(j, \tau j)}_{jm,  jm}(g)}{j^k}$ is absolute convergent and therefore convergent for all $g \in SL(2,C), \; \tau \in C, j, m \in Z$. \\[2ex]

$\square$

\section{SU(1,1)-Y-Map}
\label{sec:SU(1,1)-Y-Map}
In order to define SU(1,1)-Y-Map similar to SU(2)-Y-Map, we need two components: the Fourier coefficients of the function $\phi(v), v \in SU(1,1)$, and the matrix coefficients of the function $\psi(g), g \in SL(2,C)$. However, these matrix coefficients should be from the basis of the functions on $SU(1,1)$, rather than on $SU(2)$. Fortunately such basis exists. In all known facts of $SU(1,1)$ harmonic analysis until the SU(1,1)-Y-Map definition we will follow $\cite{Ruhl}$. \\[2ex]
Groups $SU(1,1)$ and $SL(2,R)$ are isomoprhic. Consider an element $a$ of the group $SL(2,R)$. It can be decomposed as :
\begin{equation}
a = u_1 d u_2
\end{equation}
, where $d$ =
  $\begin{bmatrix}
e^{\eta/2} & 0   \\
    0 &  e^{-\eta/2 }
  \end{bmatrix}, \; \eta \ge 0$\\[2ex]
, while $u_1$ and $u_2$ are $SU(2)$ rotation matrices of the form:\\[2ex]

$u$ =
  $\begin{bmatrix}
\cos(\psi/2) & -\sin(\psi/2)  \\
\sin(\psi/2) & \cos(\psi/2)
  \end{bmatrix}$\\[2ex]
The function $x(v)$ on $SU(1,1)$ is called  $q_1, q_2$ bi-covariant if
\begin{equation}
v = e^{i/2\psi_1 \sigma_3} e^{1/2 \eta \sigma_2} e^{i/2\psi_2 \sigma_3}
\end{equation}
implies:
\begin{equation}
x_{q_1q_2}(v) = e^{i(q_1\psi_1 + q_2\psi_2)}x_{q_1q_2}(\eta)
\end{equation}
, where $ v \in SU(1,1), \; \eta \ge 0, \; q_1, q_2 - \mbox{half-integers}, \; \psi_1, \psi_2 \in R$ \\[2ex]
It is very useful as bi-covariant functions on $SL(2,C)$ remain bi-covariant when restricted to $SU(1,1)$
\begin{equation}
\label{bi-covariant}
x_{j_1 q_1 j_2 q_2}(v) = e^{i(q_1\psi_1 + q_2\psi_2)}x_{j_1 q_1 j_2 q_2}(\eta)
\end{equation}
, where $j_1$, $j_2$, $q_1$, $q_2$ -half-integers.\\[2ex]
What important is that the functions $x(v)$ on $SU(1,1)$ can be expanded into the sum of
the bi-covariant functions on $SU(1,1)$, $\cite{Ruhl}$ (pages 128, 206):
\begin{equation}
\label{bicovariantexpansion}
x(v) = \sum\limits_{j_1q_1j_2q_2} (2j_1 + 1 )(2j_2 + 1)x^{q_1q_2}_{j_1q_1j_2q_2}(v), \;\; v \in SU(1,1)
\end{equation}
, where  $x(v)$ a) posses derivatives of all orders, b) has compact support, and c) $x_n(a)$ is a null sequence for $|a| > N$, i.e converges uniformly to zero,  \\[2ex]
and $x^{q_1q_2}_{j_1q_1j_2q_2}(v)$  are bi-covariant functions obtained from $x(v)$ in the following manner:
\begin{equation}
\label{bi-covariant}
x^{q_3q_4}_{j_1q_1j_2q_2}(v) = \int  x(u_1^{-1} v u_2^{-1}) D^{j_1}_{q_1q_2}(u_1)D^{j_2}_{q_1q_2}(u_2) d\mu(u_1) d\mu(u_2)
\end{equation}
We need to look at bi-covariant functions asymptotic behavior when $j_1, j_2 \rightarrow \infty$.  The expression ($\ref{bi-covariant}$) contains two integrals, in a sense two Fourier transforms, with respect to $u_1 \in SU(2)$ and $u_2 \in SU(2)$. For such Fourier transforms with respect to $SU(2)$ variable Paley-Wiener theorem used in ($\ref{SU(2)-Fourier}$) provided the asymptotic behavior: ($\ref{asymptotic3}$).  We can use it for bi-covariant functions in ($\ref{bi-covariant}$), writing: 
\begin{equation}
\label{asymptotic4}
|x^{q_1q_2}_{j_1q_1j_2q_2}(v)| \le \frac{C_{j_1}}{{|j_1|}^k}\frac{C_{j_2}}{{|j_2|}^k}, \;\; k \in N 
\end{equation}
We will use this result below for $SU(1, 1)$ Y-Map convergence proof. \\[2ex]
First let us consider two things that we need for providing $SU(1,1)$ Y-Map definition, i.e $SL(2,C)$ principal series matrix coefficients with the basis of functions on $SU(1,1)$, and Fourier coefficients of functions on $SU(1,1)$.\\[2ex]
Regarding $SL(2,C)$ matrix coefficients in $SU(1,1)$ functions basis, it is known $\cite{Ruhl}$ that $SL(2,C)$ principal series representation can be decomposed either into the canonical $SU(2)$ basis $D^j_{(1/2)n, q}(u)$ or $SU(1,1)$ basis: $D^J_{(1/2)\tau n, q}(v)$, where $n \in Z, \tau = \pm 1, \; s \in R, s \ge 0, \;  j, q -\mbox{half-integers}, \; u \in SU(2), v \in SU(1,1)$.\\[3ex]
$SL(2,C)$ matrix coefficients with $SU(1,1)$ basis can be defined as:
\begin{equation}
p^{n, \rho, \tau}_{j_1 q_1 j_2 q_2}(v) = <j_1 \tau q_1|T^{n, \rho}_v|j_2\tau q_2>
\end{equation}
or for the bi-covariant functions:
\begin{equation}
p^{n, \rho, \tau}_{j_1 q_1 j_2 q_2}(\eta) = <j_1 \tau q_1|T^{n, \rho}_{\exp((1/2) \eta \sigma_2)}|j_2\tau q_2>
\end{equation}
,where $T^{n, \rho}_v$ is the $SL(2,C)$ group operator, where $n \in Z, \rho \in R, v \in SU(1,1)$ \\[2ex]
$SU(1,1)$ Fourier coefficients of the discrete and continuous series representations are then as follows:\\[2ex]
For Discrete Series \cite{Ruhl} page 209:
\begin{equation}
F^{+}_{q_1q_2}(J) = \frac{1}{2} \int\limits_{0}^{\infty} x_{q_1q_2}(\eta) c^{(k, +)}_{q_1q_2}(\eta) \sinh(\eta) d \eta
\end{equation}
Continues Series  \cite{Ruhl} page 210:
\begin{equation}
F_{q_1q_2}(J) = \frac{1}{2} \int\limits_{0}^{\infty} x_{q_1q_2}(\eta) d^J_{q_1q_2}(\cosh \eta) \sinh(\eta) d \eta
\end{equation}
, where $ c^{(k, +)}_{q_1q_2}(\eta)$ and $d^J_{q_1q_2}(\cosh \eta)$ are two types of $SU(1,1)$ matrix coefficients.  \\[4ex]
Given a function $\phi(v)$ on $SU(1,1)$ we define $SU(1,1)$ Y-Map similar to $SU(2)$ Y-Map, i.e as a sum of products of $\phi(v)$ Fourier transforms $F(J)$ with the $SL(2,C)$ matrix coefficients $p^{n, \rho, \tau}_{j_1 q_1 j_2 q_2}(v)$. The solutions of $SU(1,1)$ simplicity constraints provide the map between $SU(1,1)$ and $SL(2,C)$ spins, by which Fourier transforms can be contracted with the matrix coefficients. In a sense this is what lies in the core of the Y-Map. \\[2ex]
We first write two sums below including both discrete and continuous series representations in a general form:
\begin{equation}
\label{mapdef}
	\phi_{q_1q_2}(\eta) \rightarrow \psi_{q_1q_2}(\eta)= \sum \limits_{\tau = \pm 1} \sum \limits_{J \ge 0}^{\infty} F_{q_1q_2}^{+}(J) p^{J, \tau}_{j_1 q_1 j_2 q_2}(\eta) + \sum \limits_{\tau = \pm 1} \sum \limits_{J=0}^{-1/2 +i\infty}  F_{q_1q_2}^{n, \rho} (J) p^{J, \tau}_{j_1 q_1 j_2 q_2}(\eta) 
\end{equation}
,where for Discrete Series: $J = j - 1$, \;for Continues Series: $J = -1/2 + is,  \; 0 \le s \le \infty$, 
$j_1, j_2, q_1, q_2$ - half-integers, \;\; $\tau = \pm 1, \; (n, \rho)$ - SL(2,C) principal series parameters.\\[2ex]
Then we apply the simplicity constraints solutions and sum only over them to make the above sums convergent. $SU(1,1)$ simplicity constraints solution were obtained in $\cite{ConradyHnybida}$:\\[2ex]
For Discrete Series:
\begin{equation}
\label{simplicitydiscrete}
\rho = \gamma n,  \; \;   j = n/2,  \quad \quad  \gamma \in R, n \in Z
\end{equation}
or taking into account $J = j - 1$, we can rewrite it as:
\begin{equation}
\label{simplicitydiscrete1}
\rho = \gamma n,  \; \;   J = n/2 - 1
\end{equation}\\[2ex]
For Continuous Principle Series:
\begin{equation}
\label{simplicitycont}
\rho = -n/\gamma,   \; \;  s^2 + 1/4 = -J(J+1)= \rho^2/4
\end{equation}
or
\begin{equation}
\label{simplicitycont1}
\rho = -n/\gamma,   \; \;  s = \frac{1}{2} \sqrt{(n^2/\gamma^2) -1}
\end{equation}
we select the positive sign as $s \ge 0$. By using $J = -1/2 + is$, we can rewrite it as:
\begin{equation}
\label{simplicitycont2}
\rho = -n/\gamma,   \; \;  J= -1/2 + \frac{i}{2} \sqrt{(n^2/\gamma^2) -1}
\end{equation}
, where $(n, \rho)$ are $SL(2,C)$ principal series representation parameters, $n \in Z, \rho \in R$, $\gamma \in R$ is an Immirzi constant. Note that, while in $SU(2)$ case above we were able to introduce Y-Map and prove its convergence even for complex $\gamma$, in $SU(1,1)$ case we can do it only for real $\gamma$. As it is seen from the above simplicity constraints solution, real $\gamma$ provides unitary $SL(2,C)$ principal series representations as $\rho$ becomes real, while complex $\gamma$ corresponds to non-unitary $SL(2,C)$ principal series representations.\\[2ex]
By substituting ($\ref{simplicitydiscrete}$), ($\ref{simplicitydiscrete1}$), and  ($\ref{simplicitycont2}$) into ($\ref{mapdef}$) the expressions for $J$ and $\rho$ for both discrete and continuous series,  we obtain:
\begin{equation}
\label{mapdef1}
	\phi_{q_1q_2}(\eta) \rightarrow \psi_{q_1q_2}(\eta)= \sum \limits_{\tau = \pm 1}\sum \limits_{n = 0}^{\infty} F_{q_1q_2}^{+}(n) p^{(n/2 -1), \tau}_{(j_1 q_1, j_2 q_2)}(\eta) +\sum \limits_{\tau = \pm 1} \sum \limits_{n=0}^{\infty}  F_{q_1q_2}^{n, -n/\gamma} (n) p^{( -\frac{1}{2} + \frac{i}{2} \sqrt{((n^2/\gamma^2) -1)}, \tau)}_{(j_1q_1 j_2 q_2)}(\eta) 
\end{equation}
By using expansion ($\ref{bicovariantexpansion}$) for $\phi(v), v \in SU(1,1)$ into bi-covariant functions and the above Y-Map definition for the bi-covariant functions, we finally obtain a general definition for any infinitely differentiable with compact support function on $SU(1,1)$:\\[3ex]
\textbf{SU(1,1)-Y-Map Definition:}
\begin{multline}
\label{mapdef2}
\phi(v) = \sum\limits_{j_1q_1j_2q_2} (2j_1 + 1 )(2j_2 + 1)e^{i(q_1\psi_1 + q_2\psi_2)}\phi_{j_1q_1j_2q_2}(\eta) \rightarrow \\
\sum\limits_{j_1q_1j_2q_2} (2j_1 + 1 )(2j_2 + 1) e^{i(q_1\psi_1 + q_2\psi_2)} \times \\ 
\times \left( \sum \limits_{\tau = \pm 1}\sum \limits_{n = 3}^{\infty} F_{q_1q_2}^{+}(n) p^{\frac{n}{2}-1, \tau}_{(j_1q_1, j_2q_2)}(\eta) + \sum \limits_{\tau = \pm 1} \sum \limits_{n=1}^{\infty}  F_{q_1q_2}^{n, -n /\gamma} (n) p^{( -\frac{1}{2} + \frac{i}{2} \sqrt{((n^2/\gamma^2) -1)}, \tau)}_{(j_1q_1, j_2q_2)}(\eta) \right)
\end{multline}\\[4ex]
Let us prove both sums convergence.\\[2ex]
\textbf{Theorem 2 - SU(1,1)-Y-Map Existence Theorem:} \\[2ex]
The SU(1,1) Y-Map  of an infinitely differentiable function with compact support $\phi(v), v \in SU(1,1)$ to the function $\psi(g), g \in SL(2,C)$ defined as:
\begin{multline}
\label{SU(1,1)convergence}
\phi(v) = \sum\limits_{j_1q_1j_2q_2} (2j_1 + 1 )(2j_2 + 1)e^{i(q_1\psi_1 + q_2\psi_2)}\phi_{j_1q_1j_2q_2}(\eta) \rightarrow \\
\sum\limits_{j_1q_1j_2q_2} (2j_1 + 1 )(2j_2 + 1)e^{i(q_1\psi_1 + q_2\psi_2)} \times \\
\times \left( \sum \limits_{\tau = \pm 1}\sum \limits_{n = 3}^{\infty} F_{q_1q_2}^{+}(n) p^{\frac{n}{2}-1, \tau}_{(j_1q_1, j_2q_2)}(\eta) + \sum \limits_{\tau = \pm 1} \sum \limits_{n=1}^{\infty}  F_{q_1q_2}^{n, -n /\gamma} (n) p^{( -\frac{1}{2} +  \frac{i}{2} \sqrt{((n^2/\gamma^2) -1)}, \tau)}_{(j_1q_1, j_2q_2)}(\eta) \right)
\end{multline}
is convergent.\\[4ex]
\textbf{Proof:}\\[2ex]
First, we note that the functions $p^{\frac{n}{2}-1, \tau}_{(j_1 q_1 j_2 q_2)}(\eta)$ and $ p^{( -\frac{1}{2} + \frac{i}{2} \sqrt{((n^2/\gamma^2) -1)}, \tau)}_{(j_1 q_1 j_2 q_2)}(\eta)$ are bounded, since the values of $J$, which we obtained from the simplicity constraints were based on $(n, \rho)$ of the $SL(2,C)$ principal series, and for the principal series the following bound is true (see $\cite{Ruhl}$ page 235):\\[2ex]
\begin{equation}
\label{MatrixCoeffUpperBound}
|p^{J, \tau}_{j_1 q_1 j_2 q_2}(\eta) | < C \frac{\eta}{\sinh{\eta}}, \;\; C \in R
\end{equation}
,where for the discrete series:
\begin{equation}
J = \frac{j}{2} -1
\end{equation}
,while for continuous series:
\begin{equation}
\label{Jcontinous}
J = -\frac{1}{2} + \frac{i}{2} \sqrt{(j^2/\gamma^2) -1}
\end{equation}
Remembering that $p^{J, \tau}_{j_1 q_1 j_2 q_2}$ are bi-covariant functions depending on $j_1$ and $j_2$ spins, we can rewrite the above bound in a stronger form by using the bi-covariant function asymptotic  behavior ($\ref{asymptotic4}$):
\begin{equation}
\label{MatrixCoeffUpperBound}
|e^{i(q_1\psi_1 + q_2\psi_2)} p^{J, \tau}_{j_1 q_1 j_2 q_2}(\eta) | <  \frac{\eta}{\sinh{\eta}} \frac{C_{j_1}}{{|j_1|}^k}\frac{C_{j_2}}{{|j_2|}^k},\;\;  k \in N
\end{equation}
\\[2ex]
Secondly, by Paley-Wiener theorem the Fourier coefficients of the continuous series are polynomially bounded on lines $Re(J)= const$ for bi-covariant functions $x(a)$, which vanish for $|a| > N$ see $\cite{Ruhl}$ page 218. This is exactly our case as it follows from ($\ref{Jcontinous}$) for $j > \gamma$ and $\gamma \in R$. \\[2ex]
\begin{equation}
\label{ContinuosSeriesUpperBound0}
|F(J)| \le \frac{M(N,C) \times \sup \limits_{0 \le \eta \le \infty}|D^m(\cosh(\eta) x(\eta))|}{|J(J+1)|^m}, \; \; \; m = 0, 1, 2 \mbox{...}
\end{equation}
,where $D$ is a derivative operator.\\
By substituting the simplicity constraints solution ($\ref{simplicitycont}$)  
\begin{equation}
\label{J(J+1)}
J(J+1) = -\rho^2/4 =- n^2/4\gamma^2 
\end{equation}
into ($\ref{ContinuosSeriesUpperBound0}$) we obtain:
\begin{equation}
\label{ContinuosSeriesUpperBound1}
|F(n)| \le \frac{(2\gamma)^{2m}M(N,C) \times \sup \limits_{0 \le \eta \le \infty}|D^m(\cosh(\eta) x(\eta))|}{|n|^{2m}}, \; \; \; m = 0, 1, 2 \mbox{...}
\end{equation}
We rewrite ($\ref{ContinuosSeriesUpperBound1}$) by introducing the following notation: 
\begin{equation}
\label{Notation1}
c_m =(2 \gamma)^{2m} M(N,C) \times \sup \limits_{0 \le \eta \le \infty}|D^m(\cosh(\eta) x(\eta))|
\end{equation}
as:
\begin{equation}
\label{ContinuosSeriesUpperBound}
|F(n)| \le = \frac{c_m}{|n|^{2m}} , \; \; \; j \in Z, m = 0, 1, 2 \mbox{...},\mbox{and} \; c_m = const
\end{equation}
For discrete series Paley-Wiener theorem provides a similar result: for any $m = 0, 1, 2, \mbox{...}$ $\cite{Ruhl}$ page 218:
\begin{equation}
\label{DiscreteSeriesUpperBound}
|F^{+}(J)| \le \frac{d_m}{|J(J+1)|^m}, \; \; \; m = 0, 1, 2 \mbox{...} \mbox{and} \; d_m = const
\end{equation}
By remembering that for discrete series $J = n/2 - 1$, we obtain:
\begin{equation}
\label{DiscreteSeriesUpperBound1}
|F^{+}(n)| \le \frac{d_m}{|n(n-2)|^m}, \; j \in Z, \; \; \; m = 0, 1, 2 \mbox{...}, \mbox{and} \; d_m = const
\end{equation}
Thus, we have received the upper estimates for both discrete and continuous series Fourier coefficients. 
Substituting all upper estimates: ($\ref{MatrixCoeffUpperBound}$), ($\ref{ContinuosSeriesUpperBound}$) and ($\ref{DiscreteSeriesUpperBound1}$) into SU(1,1) Y-Map definition ($\ref{mapdef2}$) we obtain:
\begin{multline}
\label{Estimate1}
\sum\limits_{j_1q_1j_2q_2} (2j_1 + 1 )(2j_2 + 1)e^{i(q_1\psi_1 + q_2\psi_2)} \times \\
\times \left( \sum \limits_{\tau = \pm 1}\sum \limits_{n = 3}^{\infty} F_{q_1q_2}^{+}(n) p^{\frac{n}{2}-1, \tau}_{(j_1q_1, j_2q_2)}(\eta) + \sum \limits_{\tau = \pm 1} \sum \limits_{n=1}^{\infty}  F_{q_1q_2}^{n, -n /\gamma} (n) p^{( -\frac{1}{2} +  \frac{i}{2} \sqrt{((n^2/\gamma^2) -1)}, \tau)}_{(j_1q_1, j_2q_2)}(\eta) \right) \le \\
\le \sum\limits_{j_1j_2} (2j_1 + 1 )(2j_2 + 1)\left(\sum \limits_{n = 3}^{\infty}  \frac{d_m}{|n(n-2)|^{m}}  \frac{2\eta}{\sinh{\eta}} \frac{C_{j_1}}{{|j_1|}^k}\frac{C_{j_2}}{{|j_2|}^k} \;\; + \;\;  \sum \limits_{n = 1}^{\infty} \frac{c_m}{|n|^{2m}} \frac{2\eta}{\sinh{\eta}} \frac{C_{j_1}}{{|j_1|}^k}\frac{C_{j_2}}{{|j_2|}^k}\right) =\\
\sum\limits_{j_1j_2} (2j_1 + 1 )(2j_2 + 1)\frac{\eta}{\sinh{2\eta}} \frac{C_{j_1}}{{|j_1|}^k}\frac{C_{j_2}}{{|j_2|}^k}  \left( \sum \limits_{n = 3}^{\infty}  \frac{d_m}{|n(n-2)|^{m}}\;\; + \;\;  \sum \limits_{n = 1}^{\infty} \frac{c_m}{|n|^{2m}} \right)
\end{multline}
, where $k > 0$.\\[2ex] 
The second sum in parenthesis in line three is convergent  for any $m \ge 1$ , since the sum value is a Riemann zeta function, while the first sum can be written as 
\begin{multline}
 \sum \limits_{n = 3}^{\infty}  \frac{1}{|n(n-2)|^{m}} = \sum \limits_{n = 3}^{\infty} \left| \frac{1}{n-2} - \frac{1}{n}\right|^m \le 
 \sum \limits_{n = 3}^{\infty} \left| \frac{1}{n-2} \right|^m +  \sum \limits_{n = 3}^{\infty} \left| \frac{1}{n} \right|^m
\end{multline}
,where we used the triangle (Minkowsky) inequality. Both sums on the r.h.s are convergent as they are Riemann zeta functions. Therefore the sum on the l.h.s. is also convergent.\\[2ex]
What remains to prove is the convergence of the first sum with respect to $j_1$ and $j_2$ in ($\ref{Estimate1}$). By using Schwarz-Cauchy-Bunyakovsky inequality, we rewrite the last line of ($\ref{Estimate1}$) as:
\begin{multline}
\sum\limits_{j_1j_2} (2j_1 + 1 )(2j_2 + 1)\frac{\eta}{\sinh{2\eta}} \frac{C_{j_1}}{{|j_1|}^k}\frac{C_{j_2}}{{|j_2|}^k}  \left( \sum \limits_{n = 3}^{\infty}  \frac{d_m}{|n(n-2)|^{m}}\;\; + \;\;  \sum \limits_{n = 1}^{\infty} \frac{c_m}{|n|^{2m}} \right) \le \\
\le  \frac{\eta}{\sinh{2\eta}}  \left( \sum \limits_{n = 3}^{\infty} \frac{d_m}{|n(n-2)|^{m}}\;\; + \;\;  \sum \limits_{n = 1}^{\infty} \frac{c_m}{|n|^{2m}} \right)\sum\limits_{j_1j_2} (2j_1 + 1 )(2j_2 + 1)  \frac{C_{j_1}}{{|j_1|}^k}\frac{C_{j_2}}{{|j_2|}^k} \le \\
\le   \frac{\eta}{\sinh{2\eta}}  \left( \sum \limits_{n = 3}^{\infty} \frac{d_m}{|n(n-2)|^{m}}\;\; + \;\;  \sum \limits_{n = 1}^{\infty} \frac{c_m}{|n|^{2m}} \right){\left ( \sum\limits_{j_1}  \frac{C_{j_1}(2j_1+1)}{{|j_1|}^k} \right )}^{1/2}  \times {\left ( \sum\limits_{j_2}  \frac{C_{j_2}(2j_2+1)}{{|j_2|}^k} \right )}^{1/2}
\end{multline}
The last two sums in in the last line are convergent for $k \ge 2$, since they are Riemann Zeta functions. This completes the proof of the SU(1,1) Y-Map convergence.
\\[2ex]
$\square$\\[2ex]

\section{ Discussion }
\label{sec:Discussion} 
First we proved $SU(2)$ Y-Map convergence, i.e $SU(2)$ Y-Map Existence Theorem, and also demonstrated  its convergence visually by using the graph calculated by MPMath Python program. Then we defined SU(1,1)-Y-Map  from infinitely differentiable functions on $SU(1,1)$ with a compact support to functions (not necessarily square integrable) on $SL(2,C)$, by using: $SL(2,C)$ matrix coefficients in the basis of functions on $SU(1,1)$, Fourier transform coefficients in both discrete and principal series, and the solution of $SU(1,1)$ simplicity constraints. We have proved that the newly defined SU(1,1)-Y-Map is convergent. This result is not to be confused with Plancherel formula for $SU(1,1)$ or $SL(2,C)$ groups. While Plancherel  formula contains a sum and an integral over all principal series parameters, ${SU(1,1)}$ and $SU(2)$ Y-Maps on the contrary, contain only sums and no integrals, and the sums are not over all spins, but only over the selected ones provided by the simplicity constraints solution.\\[4ex]

\textbf{Acknowledgment}\\
I would like to thank Michael Bukatin for both fruitful discussions and especially for his help with MPMath Python programming for generating the graph demonstrating convergence.  \\[2ex]

\section{Appendix B Numerical Demonstration}
\label{sec: NumericalConfirmation}
The numeric calculation with MPMath Python program $\cite{mpmath}$ provides the following results for the $SL(2,C)$ matrix coefficients sum convergence:
\begin{equation}
\label{SL2CMatrix3}
\sum\limits_{j=1}^{\infty}  \sum\limits_{|m| \le j} \frac{D^{(j, \tau j)}_{jm, jm}(g)}{j^k} = \sum\limits_{j=1}^{\infty}  \sum\limits_{|m| \le j}\frac{1}{j^k}{\epsilon}^{2(m + j + 1 + \frac{i \tau j}{2})} {}_2F_1( j + 1 + \frac{i \tau j}{2}, m+j+1; 2j+2; 1-{\epsilon}^4)
\end{equation}
with $\epsilon = 0.5$, Immirzi $\tau = 0.127$, for j limits from 0 to 200. The python code and the graph are as follows:\\[2ex]
For k = 2\\[2ex]
\includegraphics[width = 150mm, height=40mm]{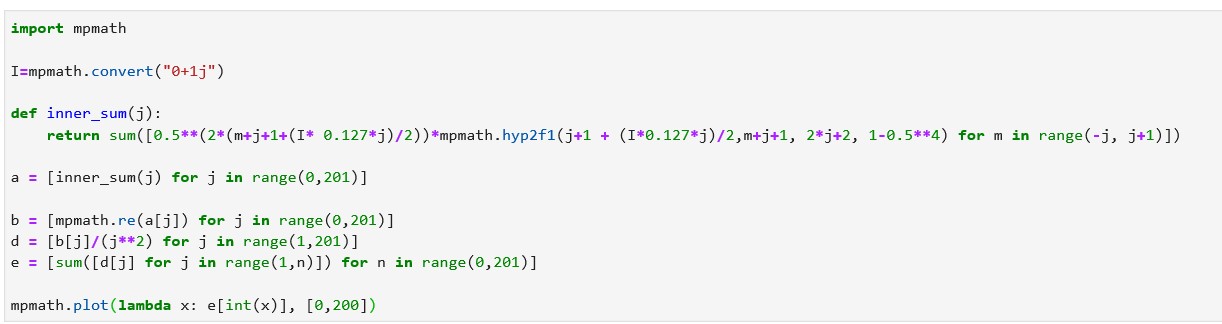}\\
\includegraphics[width = 60mm, height=40mm]{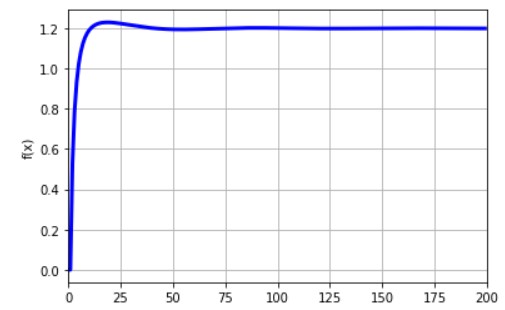}\\
For k = 6\\[2ex]
\includegraphics[width = 150mm, height=40mm]{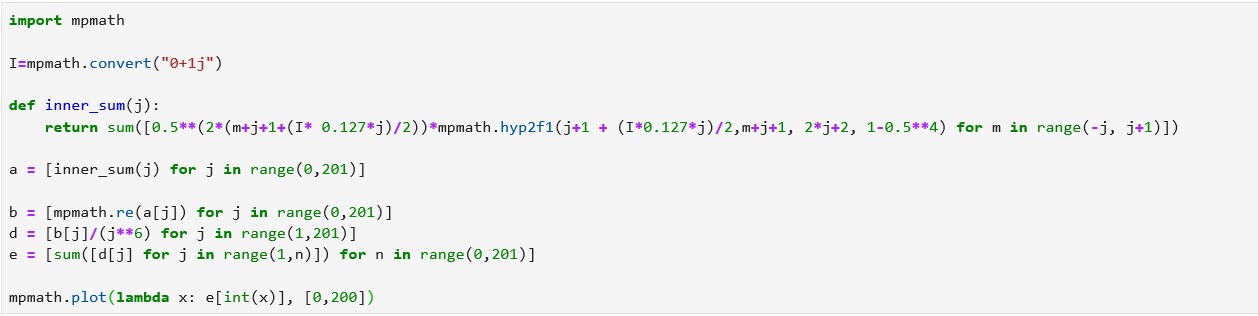}\\
\includegraphics[width = 60mm, height=40mm]{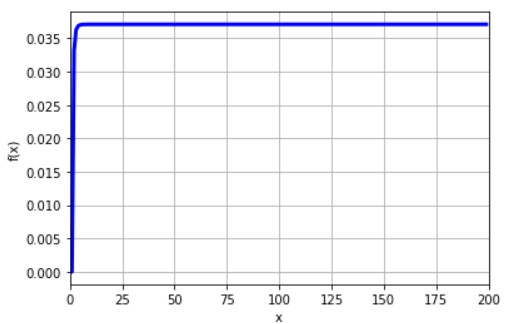}\\

\end{document}